\documentclass{article}
\usepackage{arxiv}
\usepackage[utf8]{inputenc} % allow utf-8 input
\usepackage[T1]{fontenc}    % use 8-bit T1 fonts
\usepackage{hyperref}       % hyperlinks
\usepackage{url}            % simple URL typesetting
\usepackage{booktabs}       % professional-quality tables
\usepackage{amsfonts}       % blackboard math symbols
\usepackage{nicefrac}       % compact symbols for 1/2, etc.
\usepackage{microtype}      % microtypography
\usepackage{lipsum}		% Can be removed after putting your text content
\usepackage{cite}
\usepackage{graphicx}
\usepackage{graphics}
\usepackage{xspace}
\usepackage{subfig}
\usepackage{doi}
\usepackage{caption}
\graphicspath{ {./Images/} }
\usepackage{amsmath,amssymb,amsfonts}
\usepackage{graphicx}
\usepackage{textcomp}
\usepackage{xcolor}
\usepackage[algo2e,ruled,linesnumbered]{algorithm2e}
\usepackage{algpseudocode}% http://ctan.org/pkg/algorithmicx

\title{Two Independent Teachers are Better Role Model}

\author{\href {https://orcid.org/0000-0001-9649-6351} {\includegraphics[scale=0.06]{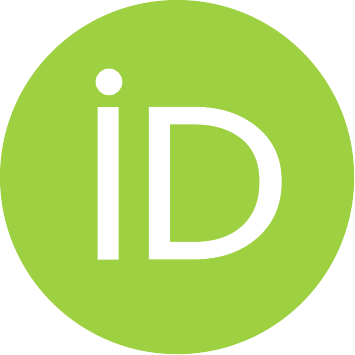}\hspace{1mm} Afifa Kahled} \\
	  School of Computer Science and Technology\\
	Huazhong University of Science and Technology\\
	  Wuhan 430074, China \\
	\texttt{afifakhaied@tju.edu.cn} \\
        \And
	\href{https://orcid.org/0000-0002-5429-2753}{\includegraphics[scale=0.06]{orcid.pdf}\hspace{1mm}Ahmed A. Mubarak} \\
	College of Applied Sciences and Educational\\
        IBB University\\
        IBB, Yemen \\
	\texttt{ahmedmubarak@ibbuniv.edu.ye} \\
	\And
	\href{https://orcid.org/0000-0001-7627-4604}{\includegraphics[scale=0.06]{orcid.pdf}\hspace{1mm}Kun He}\thanks{Afifa Kahled and Kun He are with the School of Computer Science and Technology, Huazhong University of Science and Technology, Wuhan 430074, China (E-mail: afifakhaied@tju.edu.cn; brooklet60@hust.edu.cn). Corresponding author: Kun He.\\
Ahmed A. Mubarak is with the College of Applied Sciences and Educational, IBB University, IBB, Yemen (E-mail: ahmedmubarak@ibbuniv.edu.ye).
} \\
	School of Computer Science and Technology\\
	Huazhong University of Science and Technology\\
	  Wuhan 430074, China \\
	\texttt{brooklet60@hust.edu.cn} \\
}

\renewcommand{\shorttitle}{\textit{Two Independent Teachers are Better Role Model}}

\hypersetup{
pdftitle={Two Independent Teachers are Better Role Model},
pdfsubject={q-bio.NC, q-bio.QM},
pdfauthor={Afifa Khaled, Ahmed A. Mubarak, Kun He},
pdfkeywords={Medical imaging, U-net deep network, Attention block, Teacher model, Fuse model, Brain segmentation}
}
\begin{document}
\maketitle
\begin{abstract}
Recent deep learning models have attracted substantial attention in infant brain analysis. 
These models have performed state-of-the-art performance, such as semi-supervised techniques (e.g., Temporal Ensembling, mean teacher). However, these models depend on an encoder-decoder structure with stacked local operators to gather long-range information, and the local operators limit the efficiency and effectiveness. Besides, the $MRI$ data contain different tissue properties ($TPs$) such as $T1$ and $T2$. One major limitation of these models is that they use both data as inputs to the segment process, i.e., the models are trained on the dataset once, and it requires much computational and memory requirements during inference. 
In this work, we address the above limitations by designing a new deep-learning model, called 3D-DenseUNet, which works as adaptable global aggregation blocks in down-sampling to solve the issue of spatial information loss. The self-attention module connects the down-sampling blocks to up-sampling blocks, and integrates the feature maps in three dimensions of spatial and channel, effectively improving the representation potential and discriminating ability of the model. Additionally, we propose a new method called Two Independent Teachers ($2IT$), that summarizes the model weights instead of label predictions. Each teacher model is trained on different types of brain data, $T1$ and $T2$, respectively. 
Then, a fuse model is added to improve test accuracy and enable training with fewer parameters and labels compared to the Temporal Ensembling method without modifying the network architecture. 
Empirical results demonstrate the effectiveness of the proposed method. The code is available at https://github.com/AfifaKhaled/Two-Independent-Teachers-are-Better-Role-Model.
\end{abstract}

\keywords{Medical imaging \and $U$-net deep network \and Attention block \and Teacher model \and Fuse model \and Brain segmentation.}

\section{Introduction}
\label{sec:intro}
%\input{01Intro}
%Background
Deep learning is on the horizon with the promise to enhance the segmentation of medical images due to the tedious and expensive annotating images~\cite{b1,b27}. Numerous studies have been conducted on various deep learning models that have accomplished a wide variety of tasks~\cite{b26,b28}, including brain segmentation, tumor segmentation, etc. Many of these tasks have stringent quality of result requirements. Consequently, advanced deep learning models have wide applications in brain segmentation. 

In line with that, semi-supervised techniques are undergoing a paradigm shift towards adding auxiliary tasks to improve the segmentation result and to alleviate problems in brain segmentation, particularly, formed from the $U$-net structure, as an effective alternative to other deep learning models, as $U$-net models can learn both global information and local information. 
%Furthermore, 
%However, we survey the recent $U$-net-based methods and observe some limitations. 
Though effective, we observe some limitations on recent $U$-net-based methods.

First, most of them perform segmentation on both tissue properties, called $T1$ and $T2$, based on the fact that merging $T1$ and $T2$ images together could capture more tissue information and also make segmentation an easy process. Since the differences in image contrast in MRI types, the $T1$-weighted $MRI$ enhances the fatty tissue's signal and suppresses the water's signal. In contrast, $T2$-weighted $MRI$ improves the signal of the water. In addition, the volumes of $3D$ medical images are normally too large to fit on graphics processing unit ($GPU$) memory at their native resolution, and the number of model parameters will be large, requiring enormous computation during training. Accordingly, analysis of the information provided by these modalities induces a loss of local information conducting to extreme and redundant low-level features. 

Second, by adopting convolutions and down-sampling operators, we could get local operators, scan inputs and extract local information by applying small kernels. 
%Due to the cascade way to stack the local information, the large effective kernels are the results. 
However, the cascade of stacking local information results in large effective kernels. 
The encoder generally stacks size-preserving convolutional layers, interlaced with downsampling operators, to progressively decrease the spatial sizes of feature maps. In addition, most of the models have an encoder with many stacked local operators but it suffers from a large amount of training parameters. This is because the number of feature maps generally grow doubly or even exponentially after every down-sampling operation. 
A very crucial issue in medical image segmentation is the loss of spatial information during encoding and this happens with more down-sampling, particularly when dealing with both $T1$ and $T2$ as one input. 
%Hence, the stacked local operators are the main issue in these models. 

Besides, several major challenges emerge while segmenting brain data into one or more tissues. Brain datasets are divided into $T1$ and $T2$~\cite{b10} that focus on different textures. Most existing models employ a one-stage method to segment medical data and predict the class probability and position information on both data, i.e., $T1$ and $T2$. 
Moreover, many existing popular technologies are expected to play a pivotal role in the semi-supervised learning methodology on medical images. Temporal ensembling~\cite{b2} aggregates the predictions of multiple network evaluations into an ensemble prediction, and the mean teacher method~\cite{b3} takes a supervised architecture and makes a copy of the model. In the medical image community, the mean teacher model is good for getting better predictions and has the ability to optimally exploit unlabeled data during training~\cite{b3,b29}.

This paper addresses the above limitations and challenges by proposing a 3D-DenseUNet model based on the U-net model, which focuses on loosening information loss and can leverage overall relationships between structures using multi-head attention. 
Moreover, two independent teachers are employed for deep supervision on each type of brain data. The first teacher model focuses on $T1$ and another on $T2$.  %, so each model is called a teacher model. 
Then, a fused model is adopted that takes advantage of the weights of two teacher models, leading to a decreasing number of learning parameters in the depth of network and better performance.
The performance of the proposed model is compared with the state-of-the-art models evaluated by the Dice  Coefficient metric~\cite{b8}, and empirical results demonstrate that our proposed model achieves competitive performance with fewer parameters. 

Our main contributions in terms of designing the model architecture are as follows:
\begin{itemize}
    \item We propose a 3D-DenseUNet model based on the $U$-net structure to form the down-sampling and upsampling modules, working with multi-head attention as a self-attention, which not only can effectively encode the wide context information into local features but also can mine the interdependency between channel maps and improve the feature representation of specific semantics. In addition, the skip-connections between modules to sum the multiple contextual information instead of concatenation are employed in the $U$-net model to further boost the performance. 
    \item We construct Two Independent Teacher models (2IT), with each model operating on a different data type to learn the deep feature information of each modality to reduce the information uncertainty, boundary and extract weights. 
    \item A Fuse model is built with weights updated from the two teacher modules, which could help to deeply analyze both brain data ($T1$ and $T2$), meanwhile avoid computational and memory requirements during inference and time training. 
    \item To avoid oscillation throughout the training stage of the fusion model as well as the overfitting and noising issues resulting from the summation of the teacher models' weights, we propose a function to calculate the fusing coefficient $\alpha$ value based on metric values of the model and some parameters' current epochs. 
\end{itemize}

%Paper organization
The rest of this paper is organized as follows. Section~\ref{sec:RW} presents the related work. Section~\ref{sec:methodology} presents the detailed design of our proposed model. Section~\ref{sec:results_dicussion} presents our results and discussion.  Finally, Section~\ref{sec:conclusion} concludes the paper and discusses directions for future work.  

%Related Work
\section{Related Work}
\label{sec:RW}
Deep learning has become an integral part for brain segmentation. In the realm of medical imaging, segmented images consume enormous time. It is estimated to take 9-11 hours to segment 5-20 brain images~\cite{b20}. Many proposed models based on $U$-net~\cite{b11} have achieved state-of-the-art performances in brain segmentation, 
%, including \cite{b11}. 
which adopt a serial encoding–decoding structure and use hybrid dilated convolution ($HDC$) modules. Their results are in line with a recent study in brain segmentation, which investigates concatenation between each module of two serial networks~\cite{b11}.

In addition, considerable efforts have been devoted to utilize unlabeled data in medical applications using semi-supervised techniques. For example, Zhang et al.~\cite{b5} propose a deep adversarial network ($DAN$) model that aims to attain consistently good segmentation results. Besides, another semi-supervised method~\cite{b6}  explores the learning equivalence to elastic deformations, and finds that segmentation performance improves when all training data is labeled and unlabeled data is added for training. %The authors of 
Wang et al.~\cite{b7} propose a multi-task semi-supervised segmentation model that includes a segmentation task and a regression task. A new idea is developed in \cite{b23}, where only a few number of training data is annotated. Their proposed method depends on forming a consensus prediction.

Regarding the teacher model, there are many proposed models, which try to take advantage of historical information to improve the segmentation accuracy. For example, Wang et al.~\cite{b21} propose two auxiliary tasks, and use multi-task learning  to improve performance in medical segmentation. Moreover, they propose to use uncertainty constraint to improve the feature representation. In addition, Cui et al.~\cite{b22} propose a semi-supervised method by an adapted mean teacher model, and empirical results demonstrate the effectiveness of their proposed teacher and student model. 

In contrast to previous works, we aim to 
focus on solving the problem of missing spatial information.
To this end, we propose a 3D-DenseUNet model that acts as two teacher modules to extract the weights and a fusing module that integrates the weights of two independent teacher modules instead of label predictors. Each teacher module is trained on different types of brain data ($T1$ and $T2$ respectively). Then, a fuse module takes advantage of the two modules, which improves the test accuracy and enables training with fewer labels than Temporal Ensembling \cite{b21} without modifying the network structure.  
%To validate our 3D-DenseUNet, we use the Dice coefficient metric~\cite{b8}.  
%\input{02RW}
%Methodology
\section{Methodology}
\label{sec:methodology}

The $U$-net architecture developed in 2015~\cite{b24} comprises a down-sampling as an encoder and an up-sampling as a decoder, with skip connections between them, used to segment biomedical images. 
%U-Net was developed in 2015 by Olaf Ronneberger's team~\cite{b24}.
The $U$-net model is built to integrate local and global contextual information through the encoding-decoding stages. Other recently proposed $U$-net models are based on an encoder-decoder structure with stacked local operators to gather long-range information. However, there are still two limitations in recent $U$-net models when dealing with brain segmentation tasks. First, spatial information loss is caused by down-sampling in the encoding stage. Second, the network cannot accurately locate the category of each voxel due to the loss of semantically related output and input positions caused by the network decoding phase.

\begin{figure*}
    \centering
    \subfloat{\includegraphics[width=13cm]{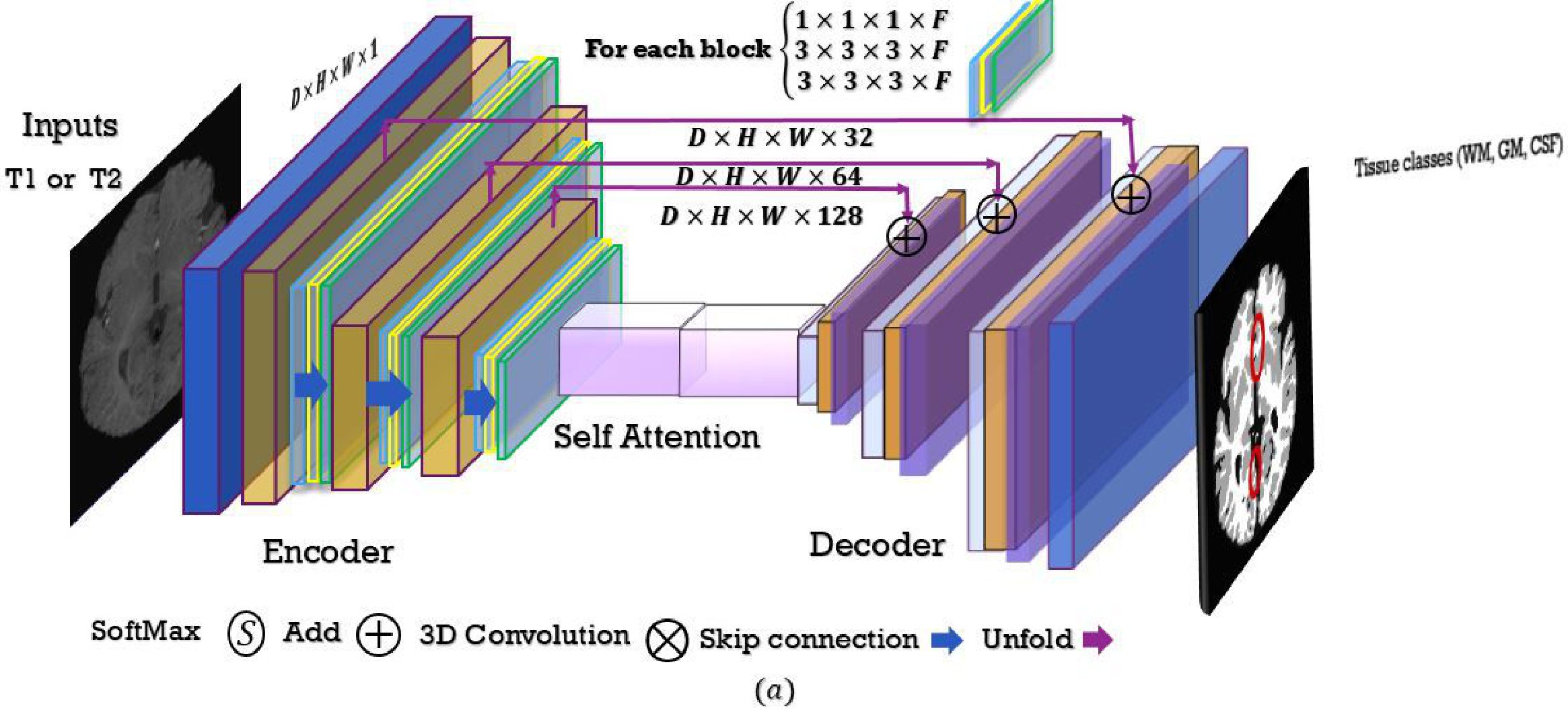}}\\
    \subfloat{\includegraphics[width=11.5cm]{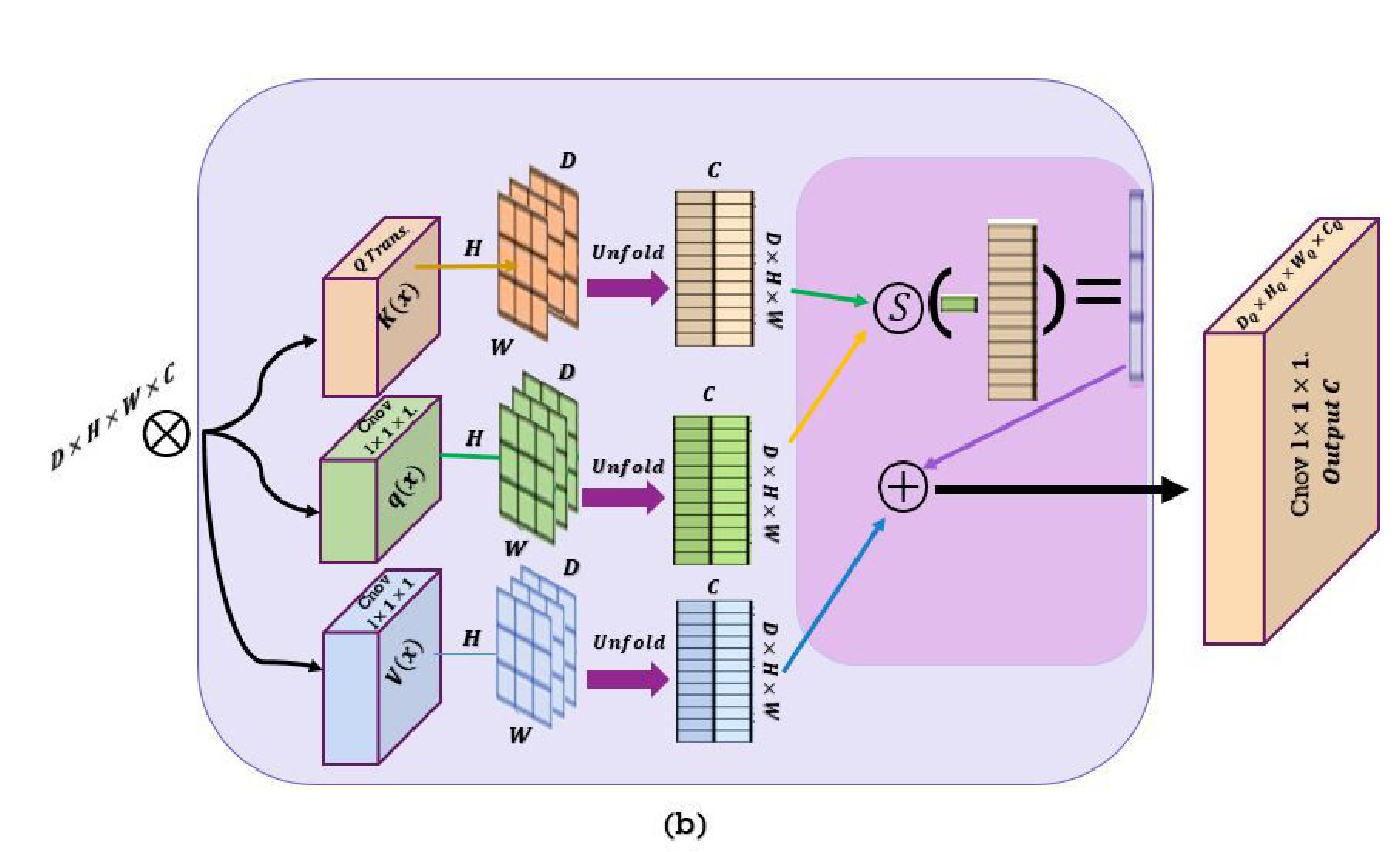}}
  \caption{Framework of the proposed model with (a) 3D-DenseUNet architecture and (b) a global attention block with 3D layers based on multi-head attention as the self-attention.}
    \label{fig:framework}
\end{figure*}

In line with that, we propose a 3D-DenseUNet model for segmenting 3D medical images. The model structure comprises a down-sampling as an encoder, which contains three blocks to build a fully residual network in each block; and an up-sampling as a decoder, which also contains three blocks to build a fully residual network in each block, with skip connections between them as long-term residual connections. In addition, a global attention block is proposed to fully use the multi-scale context features. 
Moreover, our framework comprises two independent teachers' models and a fused model, each using the 3D-DenseUNet architecture. The details of each model are presented in the following subsections. Figure. \ref{fig:framework} depicts the framework of the 3D-DenseUNet model with the global attention block employed in the proposed model.

\subsection{The Framework of 3D-DenseUNet Model}
\subsubsection{Down-Sampling}
In the proposed 3D-DenseUNet, the down-sampling module contains three blocks to build a fully residual network in each block. The residual network has three different convolutions with batch normalization and ReLU6 activation function employed before each convolutional layer.
Specifically, the first block is employed as the input layer with $3 \times 3 \times 3 $ convolution and a stride of 1, and the output block is formed all the same followed by a $1 \times 1\times 1$ convolution with a stride of 1. To solve the first issue mentioned above, the features are extracted from multiple scales using the first residual block after the sum of skip connections to avoid loss of spatial information. 

Here we use the summation instead of concatenation used in the $U$-net standard model for the following reasons. First, the summation works on the limit from increased feature maps, which contributes to reducing trainable parameters in the next layer. Second, the summation with skip connections could be regarded as long-term residual connections, which reduces the time of model training. The three different convolutions are used to build three-level features, which are combined to form the final feature for each block in down-sampling. The three different features allow down-sampling to be performed which preserves different features of the next level. %of features.
An end block then collects global information, resulting in the encoder output.

\subsubsection{Multi-head Self-attention Mechanism}
To fully use the multi-scale context features, we collect the global information of the low-level layers with the high-level feature of deep layers for final feature representations in the end block encoder. Nevertheless, a one-step sequence correlation may bypass some valuable points in the large-scale upsampling process, in which a lot of information is lost. Therefore, we need to joint feature maps of output of each apiece by attending to every position of the input semantically. We propose the attention mechanism based on the multi-head self-attention function from Transformer~\cite{b12} as shown in Figure. \ref{fig:framework} (b). The attention mechanism is adopted after the encoder stage to concatenate global information from feature maps of any size and helps to adaptively integrate information on local features and edges of the image. This block can handle the end block of down-sampling and up-sampling blocks as a block. 

In general, we have an input of 3D medical image $X^{h \times w \times d \times c}$ where $h$, $w$, and $d$ are the image shapes and $c$ the number of image channels. The end block of the down-sampling stage takes the latent cube representation $X$ as input and then computes by multi-head attention with the $1 \times 1 \times 1$ convolution and a stride of $1$. First, the attention mechanism operates to produce the query $(Q)$, key $(K)$ and value $(V)$ matrices by Eq. \ref{eq:QKV}:  
\begin{equation}
\begin{split}
Q &= Q\_Transform_{C{_K}}(X),\\
K &= Conv_{C{_K}}(X), \\
V &= Conv_{C{_V}}(X),  
\end{split}
\label{eq:QKV}
\end{equation}
where operation $Q\_Transform_{C{_K}}(\cdot)$ produces $C_K$ feature maps, and $C_K$, $C_V$ are hyper-parameters representing the dimensions of keys and values.

These matrices are decomposed in multiple heads in the second stage by parallel and independent computations. This allows a simpler stacking of multiple transformer blocks as well as identity skip connections. Therefore, the {$Unfold(\cdot)$} is employed to unfold the {$D \times H \times W \times C$} tensor into {$(D \times H \times W) \times C$} matrix, then the $V$ and $K$ dimensions are {$(H \times W \times D )$} $\times C_V$ and {$(H \times W \times D)$} $\times C_K$, respectively. 

Next, a scaled dot-product operation with SoftMax normalization between $Q$ and the transposed version of $K$ is conducted to generate the matrix of contextual attention map $A$ with dimension {$(D_Q \times H_Q \times W_Q)$} $\times$ {$(D \times H \times W)$}, which defines the similarities of given features from $Q$ concerning global elements of $K$. To calculate the aggregation of values weighted by attention weights, $A$ could be multiplied by $V$, which produces the output matrix $O$ as the following:  
\begin{equation}
O = Softmax\left(\frac{QK^T}{\sqrt{c_K}}\right)V,
\end{equation}
where $\sqrt{c_K}$ is the dimension of query $Q$ and $K$ the key-value sequence. Finally, dropout is used to avoid over-fitting and $Y$ reshapes the optimized feature maps to obtain the final output: 
\begin{equation}
Y = Conv_{CO}(Fold(O)),
\end{equation}
where $Fold(\cdot)$ is the reverse operation of $Unfold(\cdot)$ and $CO$ represents the  output dimension. So, the size of output $Y$ is $D_Q \times H_Q \times W_Q \times C_O $.

At the end of an up-sampling residual block, a $3 \times 3 \times 3$ deconvolution with a stride of 2 is employed instead of the identity residual connection, which works on taking global information into consideration, and the up-sampling block will recover more accurate details.

\subsubsection{Up-Sampling}
The up-sampling module of 3D-DenseUNet contains three blocks to build a fully residual network in each block. The residual network has three different convolutions with batch normalization and ReLU6 activation function employed before each convolutional layer. The global attention block is employed as the input layer in the up-sampling module, a $ 1 \times 1 \times 1 $ convolution with a stride of 2. Then, the output block is formed within $3 \times 3 \times 3$ convolution with a stride of 1. To solve the first issue mentioned above, the features are extracted from multiple scales using the first residual block after the sum of skip connections to avoid the loss of spatial information.
In a similar manner to the down-sampling phase, we use the summation instead of the concatenation for the limit from increased feature maps to reduce trainable parameters in the next layer. Then, we conduct a summation of skip connections as long-term residual connections, which reduces the time of training of models.  In our model for up-sampling, the $Q\_Transform_C(\cdot)$ in the global attention block is formed by a $3 \times 3 \times 3 $ deconvolution with a stride of 2 to avoid the information loss problem and keep more accurate details.  Finally, the output of the decoder block is the segmentation probability map for producing probabilities for each segmentation class using a single $\arg \max$ operation.  

\begin{figure*}[htbp]
\begin{center}  
\includegraphics[width=.8\textwidth]{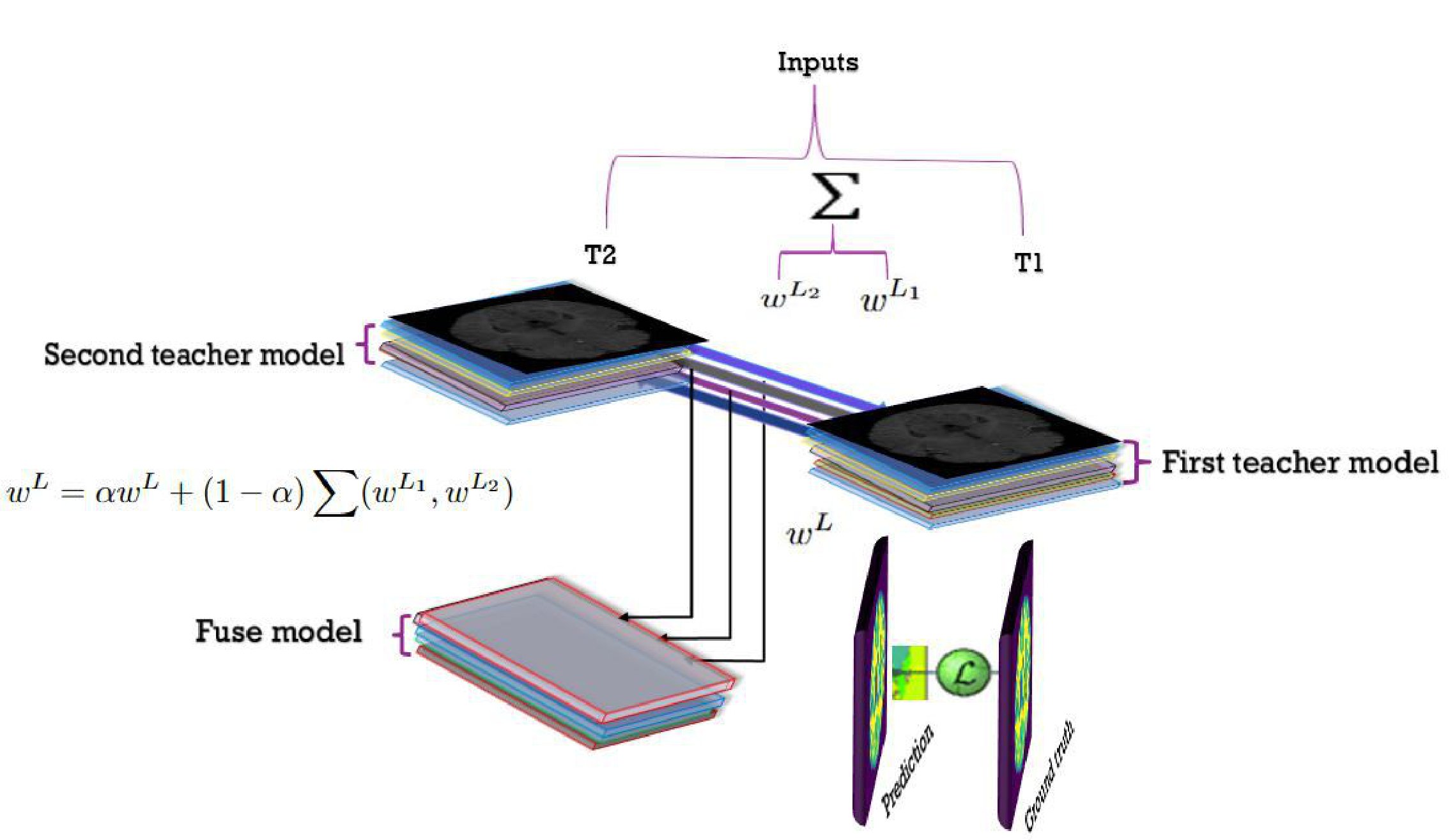}\hfill
\centering
\caption{Illustration of the fused weights from teacher models to the fuse model.}
\label{fig:fuseweights}
\end{center}
\end{figure*}

%Teacher Model
\subsection{Two-Independent-Teacher Model}
%In image data science, 
In the field of computer vision, large datasets require further analysis and extraction of features by employing deep learning models to achieve desired performance. Therefore, the models may need more training time and hyperparameters with increasing network depth, which may simply lead to degradation and is prone to over-fitting.  
Also, the resolution and image size of input affects the prediction accuracy, particularly for images in magnetic resonance $MRI$ types that share both $T1$ and $T2$ data. 
The $T1$-weighted images are produced by using short Time to Echo ($TE$) and Repetition Time ($TR$). The contrast and brightness of the images are predominately determined by $T1$ properties of tissue. Conversely, the $T2$-weighted images are produced using longer $TE$ and $TR$ times. In these images, the contrast and brightness are predominately determined by the $T2$ properties of tissue. Moreover, $T1$- and $T2$-weighted images can be easily differentiated by looking the cerebrospinal fluid ($CSF$). $CSF$ is dark on $T1$-weighted imaging and bright on $T2$-weighted imaging. Besides, $T1$-weighted $MRI$ enhances the signal of the fatty tissue and suppresses the signal of the water, while $T2$-weighted $MRI$ enhances the signal of the water.

In order to more precisely distinguish cerebral structures in neuroimaging studies and enhance tissue contrast, it may be the case that training the model with individual $T1$ or $T2$ can be an optimal way than combining $T1$ and $T2$ modalities. %On the contrary, 
However, most semi-supervised learning models consider all information these modalities provide for $MRI$ image analysis and diagnosis. Since the segmentation cost is undefined for unlabeled examples, the noise regularization by itself does not assist in semi-supervised learning~\cite{b13}. Some recent studies use noise, and then use a consistency cost between the two predictions. These studies assume a dual role as a student (it learns as before), and a teacher (it generates targets). 

%On the other hand, 
There exist at least two methods to enhance the target results. One is to select the perturbation of the representations carefully instead of barely applying additive or multiplicative noise~\cite{b14}. %Another method 
The other is to select the teacher model carefully instead of ensembling models~\cite{b3}. 
To the best of our understanding, these two approaches are consistent, and their combination may %give optimal 
yield better results.

In line with that, We take the second approach and show that it provides significant benefits. Our objective is to build two models, denoted as $TM1$ and $TM2$, as independent teachers. To adopt every model to work on different data ($TM1$ on $T1$-Weight and $TM2$ on $T2$-Weight), several changes must be made. At the first training step for each model, for the initial weights of the model, we use the strategy proposed in \cite{b15} which results in a rapid convergence of the model structure based on specific data characteristics. 
%The output of both models is averaging models' weight instead of predictions.

\subsection{The Fuse Model}
Based on the structure of the teacher model proposed by \cite{b3}, which used averages consecutive student models instead of predictions without additional training, and also to overwhelm the limitations of Temporal Ensembling~\cite{b23}. The authors used an exponential moving average ($EMA$) to calculate the averages of weights of the student model rather than sharing the weights. Their method is resulting better test accuracy and the ability to handle massive datasets. 

In keeping with that idea, we propose a Fuse model that benefits from the output of teacher models that are weights of model layers over the training steps. We can use this during training to construct a more accurate model instead of using the final weights directly according to Eq. \ref{eq:weight}. Therefore the fuse model updates weights by fusing the weight from the two independent teacher models per iteration as shown in Figure. \ref{fig:fuseweights}. Besides, through the practice of experiment, we notice that there are two advantages when using the group of independent teachers. First, summation does increase the feature maps of $T1$ and $T2$. Second, the summation of the current weight can be considered as more information, which is known to be capable of increasing the performance of the proposed model~\cite{b16}. Since the labeled data is usually very difficult to acquire, this merging weights from more than one model can open the door for a new method to improve the performance of segmentation without increasing the labeled data. 
Moreover, the main idea of our proposed model is to update each parameter's weight of the fuse model by computing the summation of weights of the same parameter in independent teacher models plus the parameter's weight multiply $\alpha$ to reduce noise and to avoid over-fitting, as shown in Eq. \ref{eq:weight}.
\begin{equation}
    W^L =\alpha W^{L}  +( 1-\alpha) \sum(W^{L_1},W^{L_2}),
    \label{eq:weight}
\end{equation}
where $W^{L_1}$ is the weight from first teacher model and  $W^{L_2}$  is weight from second teacher model. 
% \HK{Here you may need to say why design like this. It is more like the momentum update. Any motivation or reference? Could it provide more stable results? How good it is as compared with the simple average results at the end of training? Better have experiments in the next section to show the advantage of this design.}\AF{The teacher-student model that the teacher benefits from the weights generated by the student, so  I inspired by that idea and built the idea and when applying this in the fuse model, the result is better than the first teacher model and the second teacher model This is the motivation and the advantage of this design. For the alpha  already  explained in the last section}\AF{This paragraph updated to include this information}
$\alpha$ is defined in Eq. \ref{eq:alpha}. 
Furthermore, at each iteration, the 3D-DenseUNet model tunes the model weights by utilizing $\alpha$, which can be formalized as:
\begin{equation}
    \alpha =\frac{(P+1)}{((1/L)+N)},
    \label{eq:alpha}
\end{equation}
where $P$ is the accuracy, $L$ is the loss and $N$ is the batch size for each iteration.
As accuracy and loss values depend on weights of the parameters, the $\alpha$ value resulted dynamically from these values in addition to the batch size of each iteration. A careful selection of $\alpha$ can affect the performance.
In the early phase of training the proposed model, the accuracy and loss of the model obtained have not fit well enough. Thus, to avoid losing global information, $\alpha$ takes small values and the model takes full advantage of the weights of teacher models.

On the other hand, if $\alpha$ is much large, the model relies more on the current weights, leads to slow convergence and tends to overfit with low accuracy. Therefore, $\alpha$ should be close to 1. 
In line with that, we run the model several times and notice the model gains better performance when starting with 0.01 and gradually grows along with the training until approaching 1. %reaching  close to 1. 

The cross-entropy used as a cost function to measure the average (expected) divergence between the predicted output $P$ and the ground truth $T$ being approximated over the entire domain of the input sized %\textbf{\textit{W \times H \times D \times C}}. 
$ W \times H \times D \times C$.  
Let $\theta $ denote the network parameters (i.e., convolution weights, biases and $P_i$ from the parametric rectifier units), and $Y^v_s$  the label of voxel $V$ in the %$\mathrm{s}\mathrm{-}\mathrm{th\ }$
$S$-th image segment, we optimize the cross-entropy equation as following:
\begin{equation} 
J\left(\theta \right)= - \frac{\mathrm{1\ }}{S \cdot V}\sum^S_{s=1}{\sum^V_{v=1}{\sum^C_{c=1}{\delta \left(Y^v_s=C\right) \cdot {\mathrm{log} P^v_c(X_S)\ }}}}, 
\end{equation}
where $P^v_c(X_s)$ is the SoftMax output of the network for voxel $V$ and class $C$, when the input segment is $X_s$.
The overall fuse process is presented in Algorithm \ref{alg:fuse}.

\begin{algorithm2e}
\caption{The fuse model algorithm.}
\label{alg:fuse}

 \textbf{Input:} Initial model parameters $P_{^0}$, dataset $X$, number of epochs $N$, data batch $D^i$=$({x_{^i}},{y_{^i}})$, initial weights $w$. 
 
 \textbf{Output:} Prediction for segmenting $MRI$ into a number of tissues $GM$, $WM$, and $CSF$. 
 
 \For{$ n_i  \in   N $} 
{
  $ x_i \gets X$ \\
\If{$train model$ is $TM1$}
{
 \For{ $ j \in  W^{_i} $} 
 { $ w^{L_1} \gets w^{ij} $} 
 }
 \If{ $train model$ is $TM2$}
 { 
  \For{$ j \in   W^{_i} $} 
    { $ w^{L_2} \gets w^{ij} $} 
 }
\If{ $train model$ is $Fuse$}
  {  
     \For{$ j \in   W^{_i} $} 
      {
      $ Calculate$  $ \alpha $  by Eq. ~\ref{eq:alpha} \\
       $ Update $   $W^{ij}$  by Eq. ~\ref{eq:weight} 
      }
  }
 $ loss = \mathcal{L_{CE}}(x_i,y_i) $ \\
}
\end{algorithm2e}
%\input{03Method}
%Results and Discussion
%\section{Results and Discussion}
\section{Experiments and Discussion}
\label{sec:results_dicussion}
This section studies the outstanding results of the proposed model by using metrics for evaluation, and compare our method with state-of-the-art baselines. First, the evaluation metrics and datasets are described. Then, the results of the proposed form are discussed.

\subsection{Evaluation Metric}
\subsubsection{Dice Coefficient (DC)}
\label{sec:dice_coefficient}
In the research on brain segmentation, many comparison metrics have been broadly employed to assess the accuracy of segmentation methods in terms of effectiveness and efficiency. Accordingly,  Dice Coefficient ($DC$)~\cite{b8} is a powerful and important metric to assess the accuracy between predicting the model and the ground truth of data. Many recent studies used it as a benchmark and have achieved remarkable results as evaluated by the Dice Coefficient, especially, in an exploration of gray matter ($GM$), white matter ($WM$), and cerebrospinal fluid ($CSF$). It evaluates the overlap between the reference segmentation and automated segmentation. 
This metric is defined as follows:
\begin{equation}
DC(V_{ref}, V_{auto})= \frac{2 | V_{ref} \bigcap V_{auto} |}{|V_{ref}| + |V_{auto}|}, 
\end{equation}\\
where $V_{\textrm{ref}}$ denotes for the reference segmentation and $V_{\textrm{auto}}$ denotes for the automated segmentation.
The $DC$ values are given in the range of [0, 1], where 1 denotes perfect overlap and 0 denotes total mismatch.

\subsubsection{Datasets}
\begin{figure}
\centering
    \includegraphics[width=3in,height=4.5in,clip,keepaspectratio]{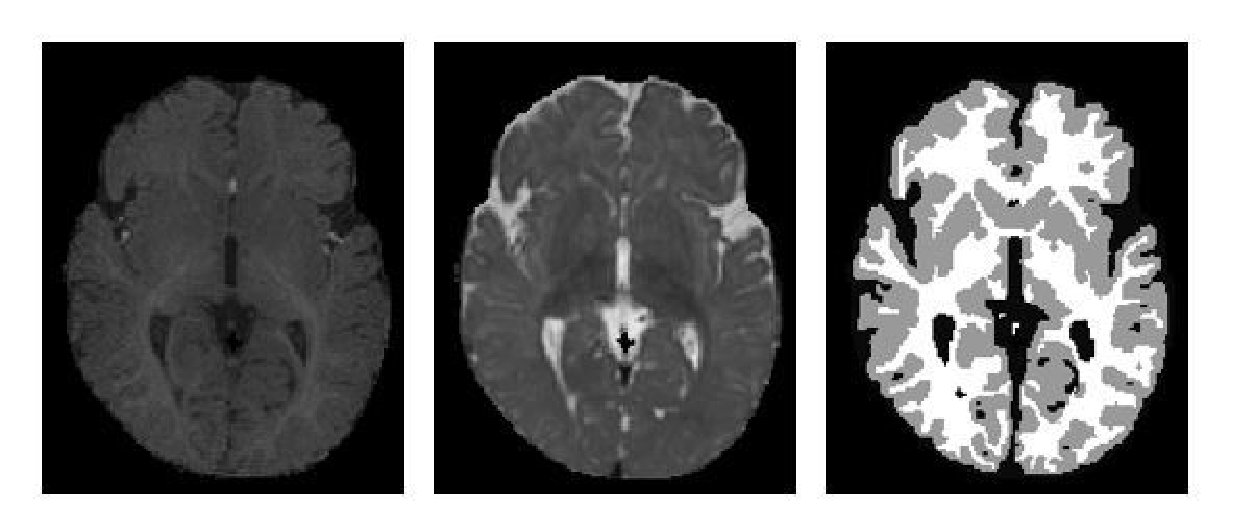}
    \caption{An example of the $MICCAI\ iSEG$ dataset ($T1$,$T2$, manual reference contour).}
    \label{fig:MICCAIiSEG}
\end{figure}

To evaluate the performance of our proposed model, we conduct experiments on $MICCAI\ iSEG$ dataset. During analysis of the $MICCAI\ iSEG$ dataset, there is a big difference in image data characteristics.
The {$MICCAI\ iSEG$} dataset was described  in \cite{b8}, that has a total of 10 images, including $T1$-1 through $T1$-10, $T2$-1 through $T2$-10, and a ground truth for the training set. And there are 13 images, $T$-11 through $T$-23, including $T$-11 through $T$-23 for test set.  
 Figure~\ref{fig:MICCAIiSEG} illustrates an example of the {$MICCAI\ iSEG$} dataset. The parameters used to create $T1$ and $T2$ are listed in Table~\ref{tab:parametersT}. 
 
\begin{table}[ht]
    \centering   
    \caption{Parameters used to generate T1 and T2.}
    \begin{tabular}{llll}
        \hline
        Data &$TR$/$TE$  & Flip angle & Resolution\\
        \hline
        T1 & 1,900/4.38 ms&7 &1$ \times $1$ \times $1\\
        \hline
        T2& 7,380/119 ms&150 
         &1.25$ \times $1.25$ \times $1.25\\
        \hline
    \end{tabular}
    \label{tab:parametersT}
\end{table}

The $MRBrainS$ is described by "https://mrbrains13.isi.uu.nl", which is an adult dataset and contains only 20 subjects. In this paper, $T1$- and $T2$-fluid-attenuated inversion recovery ($FLAIR$) images are used for segmentation. In $MRBrainS$, the number of training images are five (i.e., 2 male and 3 female). And the number of testing images are fifteen. The dataset follows the eight  tissues of segmentation of (a) peripheral cerebrospinal fluid, (b) basal ganglia, (c) cerebellum, (d) white matter lesions, (e) brainstem, (f) lateral ventricles, (g) white matter, and (h) cortical gray matter. 

\subsection{Experimental Results}
The proposed model is extensively assessed on the $3D$ multi-modal isointense infant brain tissues in $T1$- and $T2$-weighted brain $MRI$ scans task. The task is to conduct automatic segmentation of $MRI$ images into the cerebrospinal fluid ($CSF$), gray matter ($GM$) and white matter ($WM$) regions. 
In our experiment, the {$MICCAI\ iSEG$} dataset is used. For each tissue type, we first replicate the model as independent teacher models. Each model was trained on a determined data type (the first teacher model trained on $T1$ and the second on $T2$) with one sample as a validation set and the remaining nine as training sets separately. 
At each iteration, the 2000 patches of $ 32 \times 32 \times 32 $ were randomly selected as the training and validating datasets and processed in a batch size of 8. In the Fuse model process, we take patches the same size as that employed in both teacher models through the original image with a fixed overlapping step size. The size of the overlapping step should not be greater than the patch size. Thus, the third model as the fuse model benefited teacher models' weight for accurate tissue segmentation and refined tissue probability maps for every voxel in the original image for $GM$, $WM$ and $CSF$. The proposed model was trained for 5000 epochs on the training and validation datasets.

\begin{figure}
    \centering
    {\includegraphics[width=3.3in]{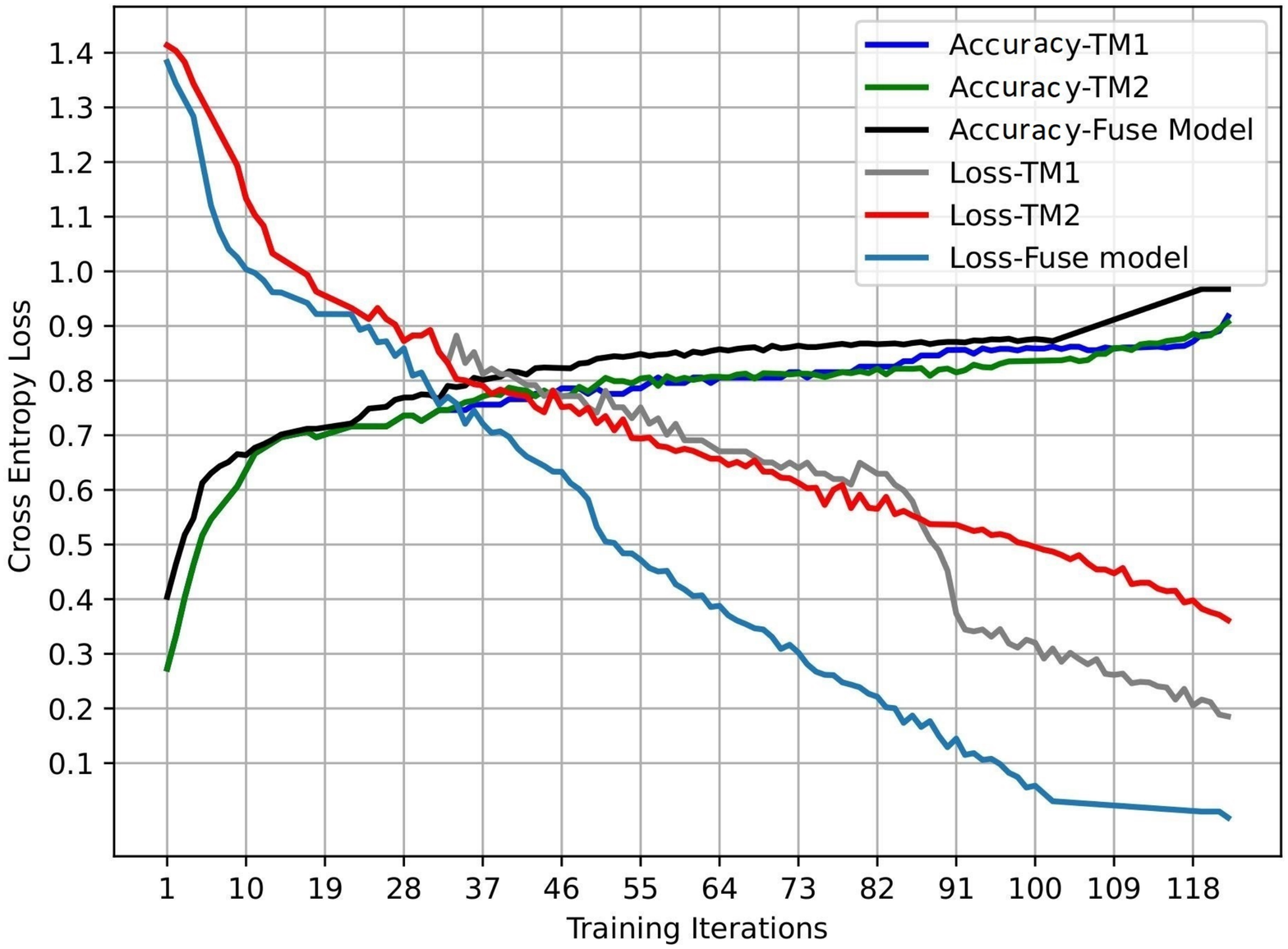}}
    \caption{The accuracy and loss validation for propose model at $ MICCAI \ iSEG$ dataset. }
    \label{fig:Segmentation accuracy}
\end{figure}

We can see that, Figure. \ref{fig:Segmentation accuracy} plots the proposed model's accuracy and loss validation stage at different iterations. The figure demonstrates that the teacher models obtained bad accuracy in the early training phase, and the loss percentage was high. On the contrary, the fuse model results were better in the early training phase. With the continuation of the training and validation phase, it was noted that any improvement in the teacher models led to an improvement in the fuse model.

%To evaluate the effectiveness of our method, 
The Dice Coefficient ($DC$) is used as the evaluation metric to evaluate the accuracy of the fuse model for the validation subject. $DC$ investigates the overlap level among the segmentation space and ground label. As we notice in Table \ref{tab:segmentation_preformance_dice_coefficient_MICCAIiSEG}, the results of the proposed model were compared to advanced state-of-the-art deep learning models in terms of metrics for the segmentation of $CSF$, $WM$, and $GM$ brain tissues. Higher $DC$ values denote more extensive overlap between manual and automatic segmentation boundary. As a result, %our model outcomes for the $DC$ metric 
our model 
has equivalent values to some models in some cases and outperforms others. Particularly, our model yields the best $DC$ values for $CSF$ and $GM$ and %slightly smaller results 
the second ranked result for $WM$.
Similarly, the 13 unlabeled images were used as the test set during the testing phase. The proposed model outperforms the state-of-the-art models in segmenting for $GM$, $WM$ and $CSF$, as shown in Table \ref{tab:segmentation_preformance_dice_coefficient_MICCAIiSEG1}.
The results reveal excellent performance compared to other methods. As we can see, the proposed model produces the best $DC$ values in $GM$, equal values in $CSF$ and is slightly smaller in $WM$ brain tissues. 
% table 2
\begin{table}[ht]
\centering
\caption{Segmentation performance in Dice Coefficient ($DC$) obtained on the $MICCAI\ iSEG$ dataset for the image used as a validation set. The best performance for each tissue class is highlighted in bold.}

    \begin{tabular}{lcccl}
        \hline
        \textbf{Model} & \multicolumn{3}{c}{\textbf{Dice Coefficient (DC) Accuracy}} \\
        \cline{2-4}
         & \textbf{~~CSF~~}       & \textbf{~~~GM~~~}         & \textbf{~~~WM~~~} \\\hline
             Wang et al. \cite{b17}    &0.95 & \bfseries 0.92 & \bfseries 0.91       \\\hline
             Hoang et al. \cite{b18}   &  0.95& 0.91& \bfseries 0.91        \\\hline
             Dolz et al.  \cite{b20}   & \bfseries 0.96 & \bfseries 0.92  & 0.90     \\\hline
             Qamar et al.  \cite{b19} & \bfseries 0.96 & \bfseries 0.92  &  \bfseries 0.91               \\\hline
             First Teacher Model   &0.80     &  0.79     & 0.90                 \\\hline
            Second Teacher Model                 & 0.89 &    0.82    & 0.81  \\\hline
             \bfseries Fuse Model  & \bfseries 0.96   & \bfseries 0.92     &  0.90           \\\hline
    \end{tabular}
    \label{tab:segmentation_preformance_dice_coefficient_MICCAIiSEG}
\end{table}
\begin{table}[ht]
\centering
\caption{Segmentation performance in Dice Coefficient ($DC$) obtained on the $MICCAI\ iSEG$ dataset for the 13 unlabeled images used for the test set. The best performance for each tissue class is highlighted in bold.}
    \begin{tabular}{lcccl}
        \hline
        \textbf{Model} &\multicolumn{3}{c}{\textbf{Dice Coefficient (DC) Accuracy}} \\
        \cline{2-4}
         %& \textbf{CSF}       & \textbf{GM}         & \textbf{WM} \\\hline
         & \textbf{~~CSF~~}       & \textbf{~~~GM~~~}         & \textbf{~~~WM~~~} \\\hline         
             Wang et al. \cite{b17}   &0.95          & 0.92 &      \bfseries 0.91        \\\hline
             Hoang et al. \cite{b18}   &  0.95       &      0.91   & \bfseries 0.91           \\\hline
             Dolz et al.     \cite{b20}    & \bfseries 0.96             & 0.92          &     0.90              \\\hline
             Qamar et al.    \cite{b19}    & \bfseries 0.96               & 0.92       &   \bfseries  0.91               \\\hline
             First Teacher Model     & 0.82         &0.80         &0.89                   \\\hline
             Second Teacher Model     &     0.88        &  0.90       &0.81 \\\hline
             \bfseries Fuse Model   & \bfseries  0.96   &\bfseries 0.93          & 0.90         \\\hline
    \end{tabular}
    \label{tab:segmentation_preformance_dice_coefficient_MICCAIiSEG1}
\end{table}

\begin{figure}
    \centering
    {\includegraphics[width=3.5 in]{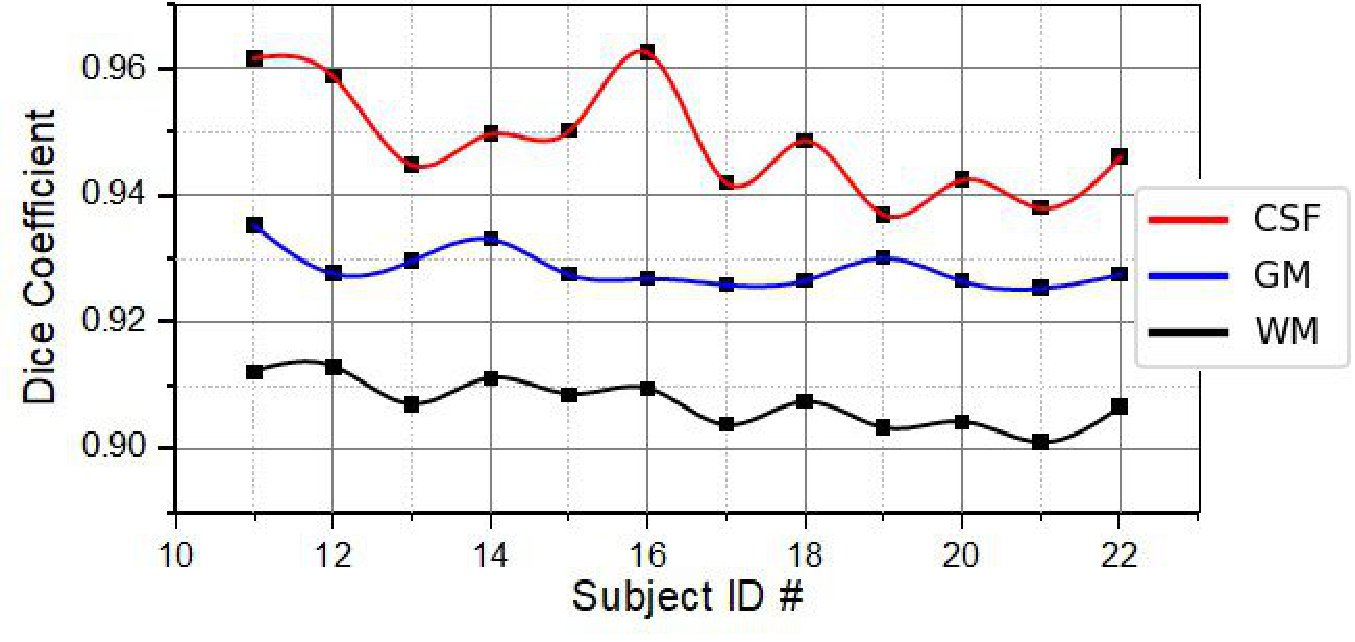}}
     \caption{Performance of the proposed model on 13 different subjects of $ MICCAI \ iSEG  $ dataset as test set.}
    \label{fig:Dice ratio}
\end{figure}

Furthermore, Figure. \ref{fig:Dice ratio} illustrates the accuracy metric values of the proposed model for the 13 subjects in the test set. As the figure shows, the $DC$ values have 
%a relatively higher variance\HK{compared with what?}
some variance in different subjects. This can be justified because providing additional training information by fused weights from two teacher models can enhance the segmentation results to significantly distinguish the edges between the white and gray matter.

In addition, we conduct an experiment on $MRBrainS$ dataset to further verify the proposed model's efficiency.
Table \ref{tab:segmentation_preformance_dice_coefficient_MICCAIiSEG2} shows the performance of our proposed model in the testing set compared with the state-of-the-art models in terms of $DC$ metric for the segmentation of $CSF$, $WM$, and $GM$ brain tissues. Our model produces good $DC$ values compared to other models, where it obtains the top results for $CSF$ and $GM$ and slightly fewer results for $WM$.

\begin{table}[htbp]
\centering
\caption{Segmentation performance (Dice Coefficient, $DC$) on the $MRBrainS$ dataset.
    The best performance for each tissue class is highlighted in bold.}
    \begin{tabular}{lccc}
        \hline
        \textbf{Model} & \multicolumn{3}{c}{\textbf{Dice Coefficient (DC) Accuracy}}\\
        \cline{2-4}
%         & \textbf{CSF}       & \textbf{GM}         & \textbf{WM} \\\hlin
         & \textbf{~~CSF~~}       & \textbf{~~~GM~~~}         & \textbf{~~~WM~~~} \\\hline 
              STH\cite{b31} & 0.83  &        0.85                 & 0.88       \\\hline   
              HyperDenseNet\cite{b30}& 0.83&   0.86     &    \bfseries  0.89     \\\hline    
             First Teacher Model    &0.39 &     0.40    &      0.50                     \\\hline
             Second Teacher Model  & 0.43 &      0.60     &      0.44   \\\hline
             \bfseries Fuse Model  & \bfseries 0.86& \bfseries0.87     &    0.87        \\\hline
    \end{tabular}
    \label{tab:segmentation_preformance_dice_coefficient_MICCAIiSEG2}
\end{table}

To be more accurate in efficiency and effectiveness, the training time was taken as a metric as shown in Table \ref{tab:execution_time} with the average training time (in hours) and standard deviation ($SD$). To study that, we compare our model with related works that reported the training time. As can be seen from the table, the average execution time of the proposed framework, which is the summation of the two teacher models and the fuse model, is lower than that of the HyperDenseNet \cite{b30}. On the contrary, it is larger than Qamar et al.\cite{b19}, despite our framework being composed of three independent models. This exemplifies that our proposed framework's architecture has fewer learned parameters. 

%Average execution time (in hours) and standard deviation ($SD$)
\begin{table}[htbp]
\centering
    \caption{Average execution time (in hours) and standard deviation ($SD$) on the $ MICCAI\ iSEG $ dataset.}
       \begin{tabular}{lccc}
        \hline
        \textbf{\centering Model~~~~~~~~~~~~~~~~~~~~~~~~~~~~~~~~~~~~} &\textbf{Time ($SD$)}  \\\hline
         HyperDenseNet\cite{b30}             &105.67 (14.7)              \\\hline
         Qamar et al. \cite{b19}             &38.00  (0.00)               \\\hline
          First Teacher Model                &36.45 (0.12)                \\\hline
          Second Teacher Model              &36.45 (0.12)                  \\\hline
          \bfseries Fuse Model              &\bfseries22.61 (0.21)          \\\hline
  \end{tabular}
  \label{tab:execution_time}
\end{table}
%A Comparison of the number of parameters.
\begin{table}[ht]
\centering
    \caption{Comparison on the number of parameters}
       \begin{tabular}{lr}
        \hline
        \textbf{\centering Model~~~~~~~~~~~~~~~~~~~~~~~~~~~~~~~~~~~~} &\textbf{Number of Parameters}  \\\hline
          Wang et al. \cite{b17}       &2,534,276~~~~              \\\hline
          HyperDenseNet\cite{b30}      &10,349,450~~~~               \\\hline
          First Teacher Model          & 2,331,160~~~~                      \\\hline
          Second Teacher Model         &2,331,160~~~~                  \\\hline
          \bfseries Fuse Model         & \bfseries2,123,211~~~~                      \\\hline
  \end{tabular}
    \label{tab:parameters}
\end{table}

Accordingly, we investigate the number of parameters used in our model. As noticed in Table \ref{tab:parameters}, our model has a number of parameters that are 20\% less than the state-of-the-art  models. The architecture of our proposed model is much deeper than all state-of-the-art models because it is composed of three independent models, which consist of 82 layers with 2.12 million learned parameters. Therefore, the proposed architecture 
is deeper compared to other existing approaches, and it achieves the 96\% accuracy. In terms of parameters, the model of Wang et al.\cite{b17} is close to ours.

\begin{figure} 
    \centering
    %\vspace{-3em}
        {\includegraphics[width=3.6in]{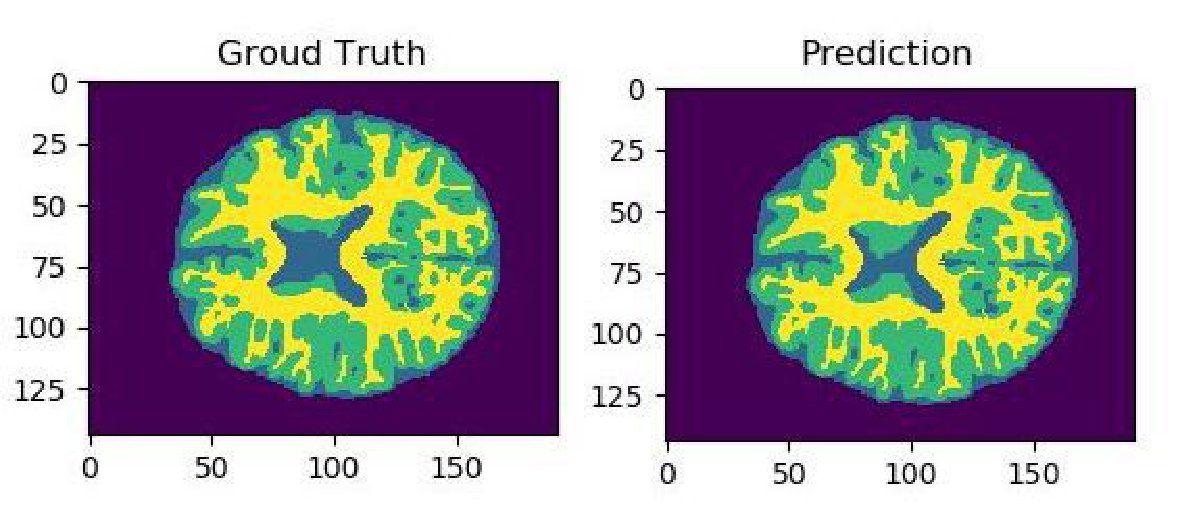}}
    \vspace{-1em}     
    \caption{Comparison on the 150-th slice results in $ MICCAI \ iSEG $ dataset. }
    \label{fig:visualization}
\end{figure}

Moreover, Figure. \ref{fig:visualization} depicts the visualization result of the proposed model on the image used as the validation set. We can see the similarity between the results achieved by the proposed model and the ground truth fairly close. Specifically, the segmentation accuracy is high at the boundary.

Overall, our results have matched the results of some models and are much better than others regarding image segmentation accuracy. Despite our model architecture being much deeper than the state-of-the-art approaches, it contains fewer parameters and takes less execution time which leads to improvement in performance comparable to the state-of-the-art approaches. 

\subsection{Ablation Study}
To evaluate the effectiveness of the component blocks used in our proposed model with the fusion method, we conduct ablation studies on the $MICCAI \ iSEG$ training and testing set. We apply different $\alpha$ values to examine what suits for our model. The two models with $\alpha$ from Eq.~\ref{eq:alpha} or with constant values for $\alpha$ are constructed and to ensure a suitable procedure. These models are trained for $5000$ epochs with a batch size of $32$ and supervised by the proposed combination loss function. Moreover, the same number of down$-$sampling and up$-$sampling blocks are used. The lowest number of alpha  starts  with  $\alpha$ =$ 0.1$ using multiple runs by changing the $\alpha$ $0.2$, $0.3$, up to $0.9$.

\begin{figure}
    \centering
    {\includegraphics[width=3.5in]{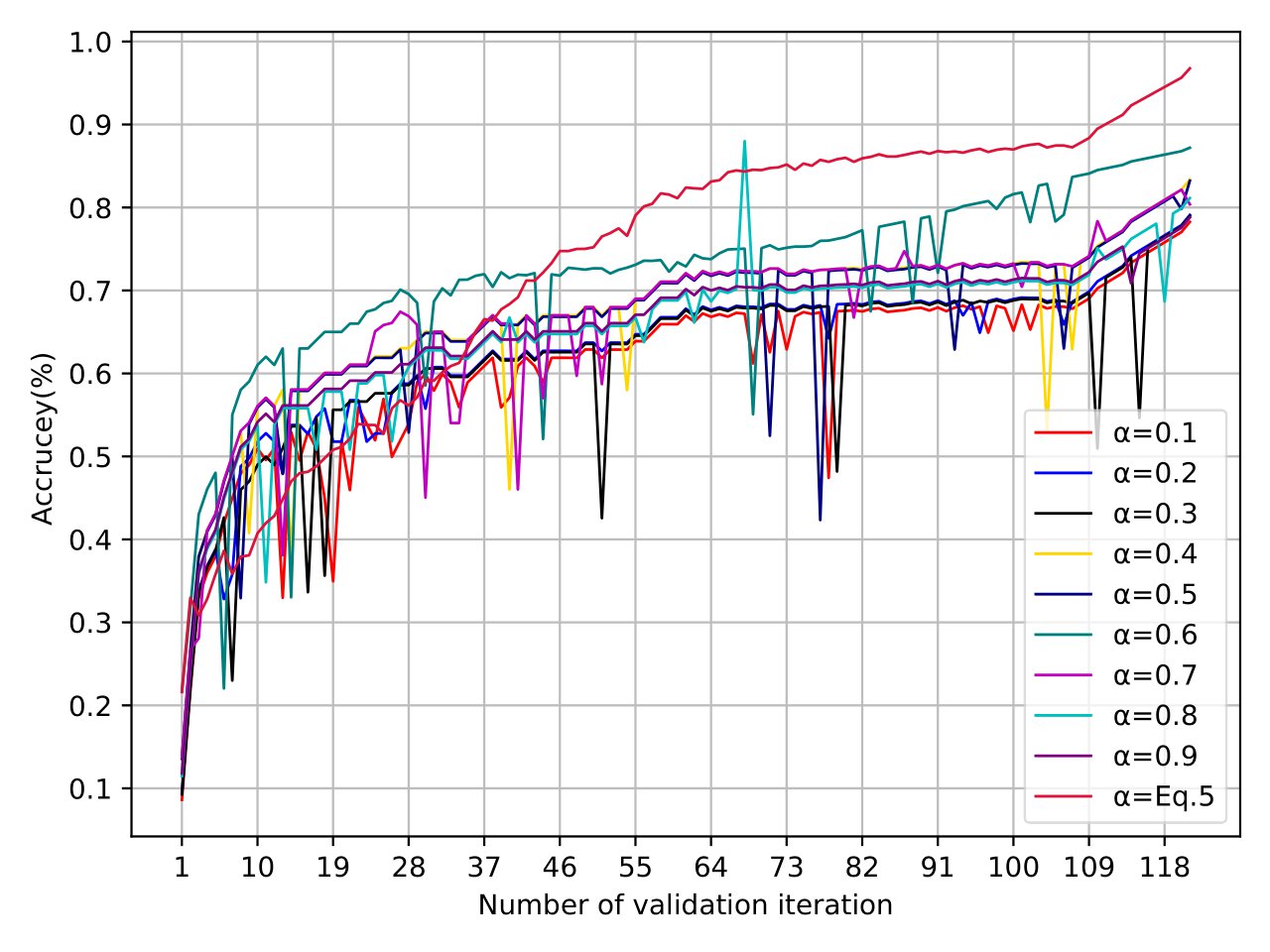}}
    \caption{The accuracy of validation dataset on $ MICCAI \ iSEG  $  with 
    various setting of $\alpha$. }
    %$\alpha$ from Eq.~\ref{eq:alpha} or with $\alpha$ = 0.1, 0.2, ..., 0.9.}
    \label{fig:alphavalue}
\end{figure}

We can see from the results that the model is improving in some values, while it is fluctuating in the values of the loss and the accuracy at some of them, especially at the minimum and maximum values which is why %the idea of 
we use Eq.~\ref{eq:alpha},  %Finally, we update $\alpha$ using Eq.~\ref{eq:alpha}. 
in that Eq.~\ref{eq:alpha} leads to better performance.
Table \ref{tab:segmentation_preformance_dice_coefficient_MICCAIiSEG3} and Figure. \ref{fig:alphavalue} report comparisons of the ablation studies on the validation set in accuracy and loss metrics. In the early training steps, the accuracy and loss of the model obtained have not fit well enough and $\alpha$ has very small values. 
In contrast, when $\alpha$ is much large, it leads to slow convergence and tends to overfit with low accuracy. Thus, the $\alpha$ values are taken according to Eq.~\ref{eq:alpha} to alleviate the fluctuations in the early stage and the overfitting issue throughout the training stage.

Also, from the point of view of the Dice metric in Table \ref{tab:segmentation_preformance_dice_coefficient_MICCAIiSEG3}, it can be observed that adopting the fusion method with $\alpha$ from  Eq.~\ref{eq:alpha} is an effective method to improve the performance of brain $MRI$ segmentation.  
These experimental results show that $\alpha$ hyper-parameter enhances the performance of the proposed model components for brain $MRI$ segmentation.

\begin{table}[ht]
\centering
\caption{Segmentation performance in Dice Coefficient ($DC$) for the $MICCAI \ iSEG$.
    The best performance for each tissue class is highlighted in bold.}
    \begin{tabular}{lccc}
        \hline
        %\multirow{\textbf{Model}} 
        \textbf{Fuse model with different $\alpha$} 
        & \multicolumn{3}{c}{\textbf{Dice Coefficient (DC) Accuracy}}\\
        \cline{2-4}
%         & \textbf{CSF}       & \textbf{GM}         & \textbf{WM} \\\hlin
         & \textbf{~~CSF~~}       & \textbf{~~~GM~~~}         & \textbf{~~~WM~~~} \\\hline
             0.1        & 0.78       & 0.74      & 0.70 \\\hline
             0.2        &   0.79     & 0.75     & 0.71  \\\hline
             0.3        &    0.79    & 0.75      &0.71   \\\hline
             0.4        &   0.83     & 0.79     & 0.76  \\\hline
             0.5        &    0.83    &  0.79    & 0.76  \\\hline
             0.6       &      0.87  &  0.83     &0.80  \\\hline
             0.7        &     0.80   &  0.75     & 0.71  \\\hline
             0.8       &    0.81    &  0.77     &  0.70 \\\hline
             0.9        &    0.79   & 0.74      &  0.70 \\\hline
             Defined in Eq.~\ref{eq:alpha} &          \bfseries 0.96     & \bfseries 0.92     & \bfseries 0.90 \\\hline
    \end{tabular}
    \label{tab:segmentation_preformance_dice_coefficient_MICCAIiSEG3}
\end{table}

\section{Conclusion}
\label{sec:conclusion}
Despite the great success of recent $U$-net models in medical image segmentation, it still suffers from some limitations. 
This work proposed a new deep-learning model, termed 3D-DenseUNet, which works as adaptable global aggregation blocks in down-sampling to solve spatial information loss. A self-attention module was connected with down/up-sampling blocks to integrate the feature maps in three spatial and channel dimensions, effectively improving its representation ability and discrimination.
Moreover, a new method called Two Independent Teachers ($2IT$) is proposed, which summarizes the model weights instead of label predictions. Each teacher model is trained on different types of brain data, $T1$ and $T2$, respectively. Then, a fused model was added to improve test accuracy and enable training with fewer parameters and labels without modifying the network architecture compared to an advanced method called Temporal Ensembling.  Empirical results showed that the proposed model outperformed other mainstream models. 
%as shown in the results section.
Overall, the results of the proposed model match the results of other models in the segmentation of brain images. This may be considered a weakness in the model, but the proposed model outperformed in terms of reducing the number of parameters and labels in addition to the implementation time.

\bibliographystyle{unsrt} 
\bibliography{Main}  %%% Uncomment this line and comment out the ``the bibliography'' section below to use the external .bib file (using bibtex).
\end{document}